\documentclass[a4paper,11pt]{article}
\usepackage{amsmath}
\begin{document}

\title{Soliton-fermion systems and stabilised vortex loops}
\author{Abhijit $\textrm{Gadde}^1$\footnote{Address after September 1 2006 :
    State University of New York at Stony Brook, USA}, 
Narendra $\textrm{Sahu}^2$ 
and Urjit A. $\textrm{Yajnik}^1$\footnote{presenter, at 17th DAE-BRNS HEP 
Symposium, IIT Kharagpur December 2006}\\
\textsl{
$1$ Indian Institute of Technology Bombay, Powai, Mumbai 400076}\\
\textsl{ $2$ Physical Research Laboratory, Navarangpura, Ahmedabad 380009}}
\date{}
\maketitle
{\flushleft
keywords : {topological solutions, zero energy modes, fractional fermion number}\\
pacs : {11.27.+d, 11.30.Er, 11.30.Fs, 98.80.Cq}
}
\section{Solitons and fermionic zero-modes}
In several self-coupled quantum field theories when treated in semi-classical limit
one obtains solitonic solutions determined by topology of the boundary conditions.
Such solutions, e.g. magnetic monopole in unified theories \cite{Hooft1974} \cite{Polyakov1974} 
or the skyrme model of hadrons have been proposed as possible non-perturbative bound states 
which remain stable due to topological quantum numbers.
Furthermore when fermions are introduced, under certain conditions one obtains
zero-energy solutions \cite{Vega1978}\cite{Jackiw1981} for the Dirac equations localised 
on the soliton. An implication of such zero-modes is induced fermion number \cite{Jackiw1977}
carried by the soliton.  

Metastable topological objects were studied in \cite{PresVil93}.
It was shown in\cite{Sahu2004} that meta-stable objects can also carry fermion zero-modes 
and this  can render a metastable object stable  
if the induced fermion number is fractional. This is true even if the fermion number is 
violated by a Majorana mass term. This result can be extended to metastable states of the 
cosmic string, in  the form of loops. These can also possess fermion zero modes\cite{Gadde06}\cite{GadYaj05} 
leading to their absolute stability. A summary of these results constitutes this conference 
contribution.

The existence of zero-modes and the possibility of fractional induced charge are well
known results for which the reader is referred to \cite{Jackiw1977}. We begin by discussing the
induced stability of metastable topological objects. 
In what follows the bosonic fields constituting the soliton are treated as classical,
and the fermion field is quantised in this background.
In such a background, the Dirac field is expanded as
\begin{equation}
\psi = c \psi_0 + \left\{ \sum_{\mathbf{\kappa}, s} a_{\mathbf{\kappa}, s} 
\chi_{\mathbf{\kappa}, s} (x) + \sum_{\textbf{k}, s}
   b_{\textbf{k}, s} u_{\textbf{k}, s} (x) + h.c. \right\}
\end{equation}
The $\psi_0$ is the solitary zero-mode assumed to exist in this example.
It is a real function and hence goes into itself under charge conjugation.
The first summation is over all the possible bound states of non-zero
frequency with real space-dependence of the form 
$\sim e^{-\mathbf{\kappa \cdot x_{\perp}}}$ 
in the transverse space directions $\textbf{x}_{\perp}$,
and the second summation is over all unbound states, which are asymptotically
plane waves.
Note that hermitian conjugates of the operators
$a_{\mathbf{\kappa}, s}$ and $b_{\textbf{k}, s}$ are included above in the
``$h.c.$'' but not that of the solitary mode operator $c$. The conjugate
operator $c^{\dag}$ appears separately in the expansion for $\bar{\psi}$.

\section{Fractional fermion number and induced stability}
For a Dirac field we have the conserved number operator satisfying
\begin{equation}
[N, \psi] = - \psi, \qquad 
[N, \psi^{\dagger}] = \psi^{\dagger}, \qquad 
\textrm{and} \quad \mathcal{C} N \mathcal{C}^{\dagger} = - N 
\end{equation}
where the last property is the charge conjugation presumed to be obeyed by the 
Lagrangian of the theory.
That these conditions are necessary though not sufficient for the existence of
fractional fermion number was shown in \cite{SudYaj86}. We next show the validity
of the same result for Majorana fermions \cite{Sahu2004}.
The Majorana condition on a fermion field requires that we
demand
\begin{equation}
  \mathcal{C} c \mathcal{C}^{\dagger} = \pm c, \qquad
\mathcal{C} c^{\dagger} \mathcal{C}^{\dagger} = \pm c^{\dagger}, \qquad
\textrm{and}\quad
\mathcal{C} N \mathcal{C}^{\dagger} = N 
\label{ccc}
\end{equation}

Unlike the Dirac case, the $c$ and $c^{\dagger}$ are not exchanged under
charge conjugation. Again we must demand the existence of a doubly degenerate
ground state with states $| - \rangle$ and $| + \rangle$ satisfying
\begin{equation}
  c| - \rangle = | + \rangle \hspace{2em} \text{\textrm{and}} \hspace{2em}
  c^{\dagger} | + \rangle = | - \rangle \label{cstates}
\end{equation}
with the simplest choice of phases. Now we find
\begin{eqnarray}
  \mathcal{C} c \mathcal{C}^{\dagger} \mathcal{C} | - \rangle & = \mathcal{C}
  | + \rangle 
\qquad \Rightarrow \hspace{2mm} \pm c ( \mathcal{C} | - \rangle) & = (
  \mathcal{C} | + \rangle)  \label{Ctransform}
\end{eqnarray}
This relation has the simplest non-trivial solution
\begin{equation}
  \mathcal{C} | - \rangle = \eta^-_M | - \rangle \hspace{2em}
  \text{\textrm{and}} \hspace{2em} \mathcal{C} | + \rangle = \eta^+_M | +
  \rangle \label{Cproperty}
\end{equation}
Now the standard fermion number operator
$ N_F = \frac{1}{2} (\psi^{\dag} \psi - \psi \psi^{\dag})$
acting on these two states gives,
\begin{equation}
  \frac{1}{2} (cc^{\dagger} - c^{\dagger} c) | \pm \rangle = \pm \frac{1}{2} |
  \pm \rangle
\end{equation}
The number operator indeed lifts the degeneracy of the two states. For $s$
number of zero modes, the ground state becomes $2^s$-fold degenerate, and the
fermion number takes values in integer steps ranging from $- s / 2$ to $+ s /
2$. For $s$ odd the values are therefore half-integral.

What is the meaning of negative fermion number of the Majorana fermions? 
We note that negative values are only induced on a soliton and the negative values
are bounded from below. We may conclude that in the presence of a soliton, the
fermionic spectrum develops a  \emph{Majorana pond} \cite{Sahu2004} at the ground state
instead of a  \emph{Dirac sea} continuing indefinitely into the negative values.

If the metastable string ruptures, each segment then can
indefinitely shrink and disappear. But then the half-integer fermion number is
not accounted for.
It was demonstrated in \cite{Sahu2004} that the rupture of the string into two
segments ending in monopoles has to be prevented by virtue of Quantum
Mechanics. This is true for both Dirac and Majorana fermions.

\section{Toroidal cosmic string }
Here we show the existence fermion zero-modes in a theroy with spontaneously 
broken local $U(1)$ \cite{Nielsen1973} with spontaneously generated majorana 
mass for fermions \cite{Jackiw1981}, but for a toroidal configuration 
\cite{Gadde06}\cite{GadYaj05}. 

We first generalise the Nielsen-Olesen solution to  toroidal 
coordinates $(v, u, \varphi)$ given by
\begin{equation}
  x = \frac{a \sinh v \cos \varphi}{\cosh v - \cos u}; \qquad 
y = \frac{a \sinh v \sin \varphi}{\cosh v - \cos u}; \qquad  
z = \frac{a \sin u}{\cosh v - \cos u}
\end{equation}
where $v$ ranges from $0$ to $\infty$, $u$ ranges from $0$ to $2 \pi$ and
$\varphi$ ranges from $0$ to $2 \pi$.
The resulting solution can be found in \cite{Gadde06}.
To solve the fermionic equations in this background, it is necessary to
introduce an ansatz which factors out the $\varphi$ dependence from upper 
and lower component respectively  of the 2-component fermion
 $\tilde{\psi}_1 = e^{- i \varphi / 2} \psi_1$ and
$\tilde{\psi}_2 = e^{i \varphi / 2} \psi_2$. This amounts to 
anti-periodic boundary condition appropriate to a fermion as we traverse 
the length of the loop. This twist is invisible for infinite length and 
the solutions reduce to those of \cite{Jackiw1981} in this limit. Now 
the equations obeyed by $\psi_1$ and $\psi_2$ are
\begin{equation}
   \left[ \begin{array}{cc}
    - [D_r + \frac{1}{2 r}] & D_z\\
    D_z & [D_r + \frac{1}{2 r}]
  \end{array} \right] \left[ \begin{array}{c}
    \psi_1\\
    \psi_2
  \end{array} \right] = g_Y \phi \left[ \begin{array}{c}
    \psi_1^{\ast}\\
    \psi_2^{\ast}
  \end{array} \right]
\end{equation}
After some analysis, the behaviour of the solutions near the loop is found 
to be of the form $e^{\pm l'v}$. Hence for $\psi$ to remain finite near the 
loop i.e. as $v \rightarrow \infty$, we find,
\begin{equation}
   0 \leq l' \leq (n - 1)
\end{equation}
This gives us a total of $n$ complex normalisable solutions, the same result as
for the infinitely long string. We are again led to the conclusion that the 
loop with odd winding number carry fractional fermion number. Further, we expect that 
by the stability argument of \cite{Sahu2004}, the loop cannot shrink indefinitely 
and disintegrate because in doing so it would have to violate fermion number by a 
fractional value.

\section{Implications for cosmology and unification }
The standard picture of cosmic string evolution is that strings can self-intersect 
or intersect with one another and in so doing produce a loop. This kind of mechanism 
is crucial to disappearance of string networks formed at cosmic phase transitions. If
they do not so intersect, they will remain and come to dominate the energy
density of the Universe.

Our considerations have shown that any unified theory which permits cosmic 
strings binding odd number of zero-modes of fermionic species \cite{SteYaj86}, 
Dirac or Majorana, must be studied carefully. The strings of odd
winding number would be stabilised and can have important implications
depending on the model. If the stabilised loops are over-abundant at the epoch of
Big  Bang Nucleosynthesis, they conflict with cosmology. Alternatively their abundance 
may be just sufficient to explain Dark Matter, being stabilised by a parity not
apparent in perturbation theory.

This work is supported by a Department of Science and Technology grant.


\begin{thebibliography}{10}
\bibitem{Hooft1974}G. `t Hooft, Nucl. Phys. \textbf{B79}, 276(1974) 

\bibitem{Polyakov1974}A. Polyakov, JETP Lett. \textbf{20}, 194(1974) 

\bibitem{Jackiw1976}R. Jackiw and C. Rebbi, Phys. Rev. \textbf{D13}, 3398(1976) 

\bibitem{Nielsen1973}H. Nielsen and P. Olesen, Nucl. Phys. \textbf{B61}, 45(1973) 

\bibitem{Vega1978}H. de Vega, Phys. Rev. \textbf{D18}, 2932(1978)

\bibitem{Jackiw1981}R. Jackiw and P. Rossi, Nucl. Phys. \textbf{B190}, 681(1981)

\bibitem{Jackiw1977}R. Jackiw, Rev. Mod. Phys. \textbf{49}, 681(1977) 

\bibitem{PresVil93}J. Preskill and A. Vilenkin, Phys. Rev. \textbf{D47}, 2324(1993)

\bibitem{Sahu2004}N. Sahu and U. A. Yajnik, Phys. Lett. \textbf{B596}, 1(2004) 

\bibitem{Gadde06} A. B. Gadde, B.Tech. Project, (2006) unpublished

\bibitem{GadYaj05}A. B. Gadde and U. A. Yajnik \textit{ArXiv:hep-ph/0511235}
 
\bibitem{SudYaj86} E. C. G. Sudarshan and U. A. Yajnik, Phys. Rev. \textbf{D33} 1830 (1986)

\bibitem{SteYaj86} See  for instance, A. Stern and U. A. Yajnik, Nucl. Phys. \textbf{B267}, 158 (1986) 
\end{thebibliography}
\end{document}